\documentclass[aps,prl,reprint,showpacs,showkeys]{revtex4-1}
\usepackage{amsfonts,amssymb,amsmath,epsfig, graphicx,yfonts}
\usepackage[utf8]{inputenc}

\makeatletter
\renewcommand*{\@fnsymbol}[1]{\ensuremath{\ifcase#1\or \ddagger\or
    \mathsection\or \mathparagraph\or \|\or **\or \dagger\dagger
    \or \ddagger\ddagger \else\@ctrerr\fi}}
\makeatother

\newcommand{\be}{\begin{equation}}
\newcommand{\ee}{\end{equation}}

\begin{document} 
\title{Melting holographic mesons by applying a magnetic field}
\date{\today}

\author{Daniel \'Avila,}
\email{davhdz06@ciencias.unam.mx}
\author{Leonardo Pati\~no,}
\email{leopj@ciencias.unam.mx}
\affiliation{Departamento de F\'isica, Facultad de Ciencias, Universidad Nacional Aut\'onoma de M\'exico, \\  A.P. 70-542, M\'exico D.F. 04510, Mexico} 

\begin{abstract}
In the present letter we use holographic methods to show that a very intense magnetic field lowers the temperature at which the mesons melt and decreases the mass gap of the spectrum along with their masses. Consequently, there is a range of temperatures for which mesons can be melted by applying a magnetic field instead of increasing the temperature. We term this effect Magnetic Meson Melting (MMM), and we are able to observe it by constructing a configuration that makes it possible to apply gauge/gravity methods to study fundamental degrees of freedom in a quark-gluon plasma subject to a magnetic field as intense as that expected in high energy collisions. This is achieved by the confection of a ten-dimensional background that is dual to the magnetized plasma and nonetheless permits the embedding of D7-branes in it. For such a background to exist, a scalar field has to be present and hence a scalar operator of dimension 2 appears in the gauge theory. We present here the details of the background and of the embedding of flavor D7-branes in it. Since our results are obtained from the gravity dual of the gauge theory, the analysis is also interesting from the gravitational perspective.

\end{abstract}

  

\maketitle
\setlength{\parskip}{3pt}

\section{Introduction and main results}

It has become increasingly accepted that an intense magnetic field is produced in high energy collisions and that understanding its effects is relevant to properly analyze experimental observations 
\cite{Skokov:2009qp,Wilde:2012wc,Basar:2012bp,Ayala:2018wux}. A tool that has proven to be very useful to study the quark-gluon plasma produced in these collisions is the gauge/gravity correspondence 
\cite{Maldacena:1997re}. In this context, adding $N_{f}$ massive flavor degrees of freedom to a gauge theory corresponds to embedding $N_{f}$ D7-branes in its gravitational dual \cite{Karch:2002sh}. The incorporation of the magnetic field in this setup was done in \cite{Filev:2007gb,Albash:2007bk}, where said field was introduced as an excitation of the D7-branes, and the latter were considered to be a probe in a fixed ten-dimensional background. The results obtained in \cite{Albash:2007bk} showed that the effect of the magnetic field was to increase the dissociation temperature of the mesons along with their masses, thus providing an holographic realization of magnetic catalysis (MC). However, it is also well known from lattice calculations \cite{Bali:2017ian} and linear sigma models \cite{Ayala:2018zat} that for some mesons, such as the neutral pion, the magnetic field has the opposite effect on both, its mass and dissociation temperature. It then becomes necessary to find a manner in the gauge/gravity correspondence to reproduce this inverse magnetic catalysis (IMC) for meson dissociation (IMC for other physical phenomena has been reported in various holographic models \cite{Preis:2010cq,Mamo:2015dea,Rodrigues:2017iqi}).

A different approach to incorporate the effects of a strong magnetic field in the correspondence was followed in \cite{DHoker:2009mmn}, where the plasma contained massless matter in the adjoint representation. Trying to implement the embedding of the flavor D7-branes in the ten-dimensional uplift of the five-dimensional geometry employed in \cite{DHoker:2009mmn} turns out to be highly complicated because the compact part of the resulting geometry warps in a way that prevents an easy identification of the right 3-cycle that the D7-brane must wrap \cite{Uriel}. Given this complication, our approach is to construct a family of solutions to ten-dimensional type IIB supergravity that accommodates a magnetic field and has a compact five-dimensional space that factors as a warped 3-sphere, a 1-cycle, and an angular coordinate $\theta$ that determines the volume of these two spaces. By keeping the 1-cycle and the $\theta$ direction perpendicular to the rest of the spacetime, we will be able to proceed as in previous approaches 
\cite{Mateos:2007vn}, and have the D7-brane naturally wrapping the 3-cycle. We will see below that the construction of such a family requires the excitation of a scalar field $\varphi$ dual to a scalar operator of dimension 2 in the gauge theory, that hence saturates the BF bound \cite{Breitenlohner:1982jf}.

In \cite{Avila:2018hsi} we adopted a five-dimensional effective perspective to study this family of backgrounds, and found a critical intensity $b_c$ for the magnetic field above which the solutions become unstable.

The principal purpose of this letter is to show how, in the holographic setup that we study, a magnetic field lowers the temperature at which the meson melting transition happens, implying that for certain temperatures such a magnetic field can be used to cause this transition. Adjacent to this result, by studying mesons that are dual to perturbations on the embedding of the brane, we find that their masses decrease with the intensity of the magnetic field, as does the mass gap of the spectrum. Thus, we provide an holographic realization of IMC for meson dissociation.

Not less important, regarding the gauge/gravity correspondence, is the presentation we do of the ten-dimensional family of backgrounds and the embedding of D7-branes into it, both of which we constructed numerically, since analytic solutions eluded our treatment. Numerical details of what we present here, along with further information about physical quantities affected by the magnetic field in this context, will be available in 
\cite{MagneticMesons}.

\section{The gravitational background}

What we need is a solution to ten-dimensional type IIB supergravity with a metric that asymptotically approaches AdS$_5\times S^5$, and accommodates a deformation that encodes the dual of a magnetic field in the gauge theory, while still permits to factor out the compact part of the metric in the way described in the introduction.

As it turns out, a general line element that allows such a solution is given by
\begin{eqnarray}
ds_{10}^2&=&\sqrt{\Delta}ds_{5}^2\nonumber\\
+\frac{1}{X\sqrt{\Delta}}&&\!\!\!\!\!\!\left(X^2\Delta d\theta^2+X^3\sin^2\theta d\phi^2+\cos^2\theta {d\Sigma_3}^2\right), \label{ds10Dft}
\end{eqnarray}
where $X=e^{\frac{1}{\sqrt{6}}\varphi(r)}$, $\Delta=X\cos^{2}\theta+X^{-2}\sin^{2}\theta$, and the line element $ds_{5}^2$ of the non-compact subspace has the form
\be
ds_{5}^{2}=\frac{dr^{2}}{U(r)}-U(r)dt^{2}+V(r)(dx^{2}+dy^{2})+W(r)dz^{2}, \label{ds5}
\ee
while the one of the 3-cycle is given by
\be
{d\Sigma_3}^2=d\vartheta_{1}^{2}+\sin^{2}\vartheta_{1}(d\vartheta_{2} +\sqrt{2}A)^{2}+\cos^2\vartheta_{1}(d\vartheta_3 +\sqrt{2}A)^{2}.\label{3cycle}
\ee

The coordinate $r$ measures a radial distance, and we expect \eqref{ds10Dft} to approach AdS$_5\times S^5$ as $r\rightarrow\infty$, making the directions $t, x, y,$ and $z,$ dual to those in which the gauge theory lives.

We see that the 1-form $A$ parametrizes an infinitesimal rotation involving a periodic direction of the compactifying manifold that, in turn, codifies the internal degrees of freedom of the dual gauge theory. If we keep $A$ and its exterior derivative in the cotangent space to the directions dual to those of the gauge theory, it will represent a U(1) vector potential that allows the introduction of the desired magnetic field in the latter. In the family of solutions that will provide the background for our current calculation we set $A=b\, x\,dy$, automatically satisfying Maxwell equations 
\cite{Avila:2018hsi} and introducing a constant magnetic field $F=b dx\wedge dy$ in the gauge theory.

Given that the scalar $\varphi$ and metric potentials $U, V,$ and $W,$ that we numerically constructed in 
\cite{Avila:2018hsi} are solutions to 5D gauged supergravity, their substitution in the expressions above guaranties for \eqref{ds10Dft} to solve the equations of motion of type IIB supergravity in 10D
\cite{Cvetic:1999xp}, just as long as the 5-form field strength proper to this theory is given by the expression also included in 
\cite{Avila:2018hsi}.

This is the family of backgrounds that we will use, of which all elements posses a regular horizon at some finite $r_h$, providing the gauge theory with a finite temperature $T=\frac{3r_h}{2\pi}$, while the non compact part of \eqref{ds10Dft} asymptotes to AdS$_5$ for large $r$. Since the 5-form does not couple to the D7-brane, it will not appear in our current calculations and will be omitted it in this letter.

\section{Embedding of the D7-brane}

To find how a D7-brane is embedded in our background we must extremize the Dirac-Born-Infeld action given by
\begin{equation}
S_{DBI}=-T_{D7}N_{f}\int d^{8}x\sqrt{-\text{det}(g_7)},
\label{DBI}
\end{equation}
where $T_{D7}$ is the tension of the D7-brane, $g_7$ the metric induced over it, and the integration is to be performed over its world volume.

An extreme of \eqref{DBI} can be consistently found at fixed $\phi$ given that \eqref{ds10Dft} does not depend on this coordinate, and because the direction that $\phi$ represents remains orthogonal to the rest of the spacetime. Notice that achieving this orthogonality is what made the introduction of $\varphi$ necessary.

Concerning the 3-cycle in \eqref{ds10Dft}, we notice that its volume depends on the position $\theta$ and, regardless of the value of $b$, this cycle becomes maximal at $\theta=0$, while for $b=0$ it reduces to $S^3$. For non-vanishing $b$, the 3-cycle gets tilted towards the five dimensional non-compact part of the spacetime in a manner that is volume preserving within the eight dimensions of this two spaces together.

We see then that a D7-brane can consistently extend along the directions of ${ds_5}^2$ and ${d\Sigma_3}^2$, and since the volume of this subspace depends solely on $\theta$ and $r$, an embedding that extremizes \eqref{DBI} can be found by setting $\phi$ to a constant and determining the right profile for $\theta(r)$. This embedding becomes supersymmetric at zero temperature when we turn off $\varphi$ and $A$.

The expressions to follow simplify significantly if they are written in terms of $\chi(r)\equiv \sin\theta(r),$ so that the line element induced on the D7-brane is
\begin{eqnarray}
ds^{2}_{D7}&=&\Delta^{\frac{1}{2}}\left[-Udt^{2}+V(dx^{2}+dy^{2})+Wdz^{2}\right. \cr
&+&\left. \frac{1-\chi^{2}+UX(\partial_{r}\chi)^{2}}{U(1-\chi^{2})}dr^{2}\right]+\frac{1-\chi^{2}}{\Delta^{\frac{1}{2}}X}d\Sigma_{3}^{2}(A),
\label{metric_D7}
\end{eqnarray}
where the wrapping factor simplifies to $\Delta=X+\chi^{2}(X^{-2}-X).$ We find the embedding by varying \eqref{DBI} with respect to $\chi$ after substitution of \eqref{metric_D7} and solving the resulting equation.

The behavior of the embedding close to the boundary is given by $\chi=\bar{M}\pi/\sqrt{2}r+...$, where $\bar{M}$ is related to the quark mass by $M_{q}=\bar{M}\sqrt{\lambda}/2$, with $\lambda$ the t'Hooft constant. For values of $\bar{M}$ not too large in comparison to the temperature, the embedding at any $r$ remains close enough to the equator, $\chi(r)=0$, and the brane falls trough the horizon as $r\rightarrow r_h$, receiving the name of black hole embedding. On the contrary, for large enough values of $\bar{M}$ the embedding stays distant from the equator and the brane does not touch the horizon, so it is refereed to as a Minkowski embedding. There is an intermediate range of $\bar{M}$ values for which there are both types of embeddings, and a thermodynamic analysis is necessary to determine which one is favored.

Once a D7-brane has been introduced in this background, open strings with both ends on it can exist. The low energy states of these strings are dual to mesons, \textit{i.e.} quark-antiquark bound states, in the gauge theory. These states are codified as excitation of the D7-brane governed by the DBI action, and their spectrum can be determined by finding the stable vibrational and U(1) perturbations of the brane. It turns out \cite{Hoyos:2006gb,Mateos:2007vn} that for Minkowski embeddings the spectrum is discrete and has a non-vanishing mass-gap, while for black hole embedding it is continuous and gapless. From this it is concluded that Minkowski embeddings are dual to a phase of the gauge theory where stable mesons exist, while black hole embeddings correspond to a different phase in which the mesons have dissociated or melted. When the ratio $\bar{M}/T$ is considerably above the value at which the system transit from one phase to the other, $\bar{M}$ is related to the mass gap of the meson spectrum \cite{Kruczenski:2003be,Arean:2006pk,Ramallo:2006et,Myers:2006qr}, so we see that the transition is governed by how this physical quantity compares to the temperature.

If the transition between embeddings is thought of at fixed $\bar{M}$ but changing temperature, we see that at low temperatures, represented by small $r_h$, the D7 does not fall through the horizon, and stable mesons exist. As the temperature increases the system gets to a point in which the brane does fall through the horizon and the mesons melt.

\section{Magnetic meson melting (MMM)}
\label{numerical}

\begin{figure}[ht!]
 \centering
 \includegraphics[width=0.45\textwidth]{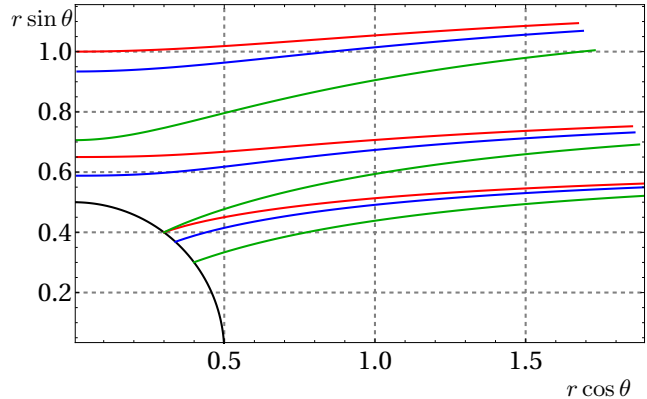}
 \put(-30,-10){$r\cos\theta$}
 \put(-240,130){$r\sin\theta$}
\caption{{\footnotesize Profiles in the $(r\cos\theta,r\sin\theta)$-plane. The circle represents the horizon. From highest to lowest bundle $\bar{M}=\lbrace .55, .38, .29\rbrace$. In each bundle, red, blue and green (top to bottom) correspond to $b/T^{2}=\lbrace0,5.45,11.04\rbrace$ respectively.}}
\label{embeddings}
\end{figure}

To visualize our main result, in Fig. \ref{embeddings} we display several profiles at the same temperature $T=\frac{3}{4\pi}$, grouped by their value for $\bar{M}$. Our first finding is that for stronger magnetic fields the brane bends closer to the horizon, lowering the value of $T/\bar{M}$ at which the mesons melt. A second observation is that there is a range for $\bar{M}$, 0.38 serving as an example, for which the intensity of the magnetic field can change the embedding from Minkowski type to black hole. This means we can keep $T/\bar{M}$ fixed and use the magnetic field to drive the transition that will happen at a certain critical intensity $b_{mmm}$. Importantly, since the bound $b_c$ for the intensity of the magnetic field is inherited from the analysis in \cite{Avila:2018hsi}, this melting can only occur for a certain range in $T/\bar{M}$.

To study this transition more carefully we compute the free energy associated to a number of embeddings of both types covering those close to the transition. The free energy is given by the product ${\cal{F}}=T\,S_{D7}$ of the temperature $T$ characterizing the background and $S_{D7}$, which is the renormalized Euclidean continuation of \eqref{DBI} evaluated on the corresponding solution $\chi(r)$.

The details of the renormalization for $S_{D7}$, that will be available in \cite{MagneticMesons}, indicate the existence of some freedom in choosing the renormalization scheme. Since all the qualitative results are scheme independent, in this letter we work in a fixed scheme of which the particulars are not relevant, and leave other instanses for \cite{MagneticMesons}.

\begin{figure}[ht!]
 \centering
 \includegraphics[width=0.47\textwidth]{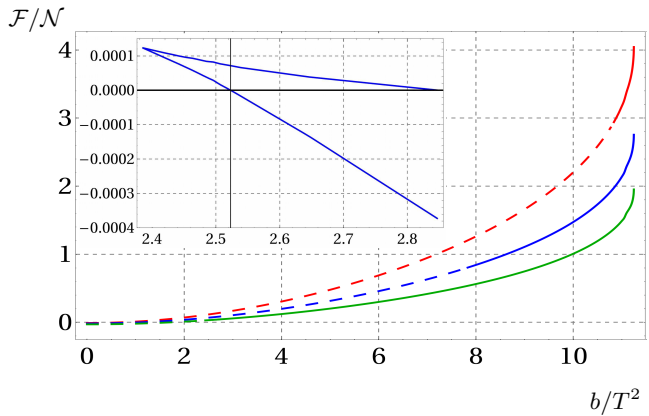}
 \put(-30,-10){$b/T^{2}$}
 \put(-250,135){$\mathcal{F}/\mathcal{N}$}
\caption{\footnotesize{$\mathcal{F}/\mathcal{N}$ as function of $b/T^{2}$. Red, blue and green (bottom to top) correspond to $\bar{M}/T=\lbrace {1.81,1.53,1.33} \rbrace$ respectively. Dashed segments correspond to Minkowski embeddings, while continuous segments correspond to black hole embeddings. $\mathcal{N}=T_{D7}N_{f}\text{vol}(S^{3})\text{vol}(x)\frac{\pi^{3}T^{4}}{4}$.}}
\label{FreePlot}
\end{figure}

In Fig. \ref{FreePlot} we fix $T=\frac{3}{4\pi}$ and show the behavior of ${\cal{F}}$ as a function of $b/T^2$ for three values of $\bar{M}/T$ that permit both phases for different intensities of the magnetic field. These plots explicitly show that the phase transition can be driven by the magnetic field and demonstrate our main result. The inset shows the difference in free energy between the black hole and Minkowski embeddings close to the melting point of the mesons for the case $\bar{M}/T=1.53$. In this example $b_{mmm}/T^{2}=2.52$.

Extending the thermodynamic analysis of the transition requires to find the temperature dependence of the free energy at fixed magnetic field, the entropy density computed from it, and the specific heat $C_b$. This will be presented in 
\cite{MagneticMesons}.

\section{Meson spectrum}

The spectrum of the mesons in the gauge theory can be determined by finding the stable normal modes of the Minkowski embeddings 
\cite{Karch:2002xe,Kruczenski:2003be,Arean:2006pk,Ramallo:2006et,Myers:2006qr} for excitations of either vibrational modes or those of the world volume U(1) field. To demonstrate the effect that the magnetic field has over the meson spectrum, in this letter we will only study those that correspond to perturbations of the embedding in directions perpendicular to it and leave other sectors to be presented in \cite{MagneticMesons}.

A general excitation of this kind can be implemented by writing $\chi(X)=\chi_0(r)+\delta\,\chi_1(X)$ and $\phi(X)=\phi_0+\delta\,\phi_1(X)$, where the naughted functions are the solutions discussed earlier and the perturbations, denoted by a subindex 1, can depend on any coordinate of the ten-dimensional space. The equations of motion for these perturbations show that $\chi_1(X)$ and $\phi_1(X)$ decouple from each another, and furthermore, their dependence on the 3-cycle coordinates, on $r$, and on the gauge theory directions, can be factored. Once reduced over the 3-cycle, the dependence of the perturbations on its coordinates is related to the $r$-charges of the states in a Kaluza-Klein tower. For concreteness we will focus on perturbations $\chi_1(X)$ that do not depend on the coordinates of the 3-cycle and leave other examples for 
\cite{MagneticMesons}.

Even if the magnetic field makes our gauge theory not isotropic, it remains invariant under translations, so we can write $\chi_1(X)=e^{i\omega t-k_\mu x^\mu}\chi_r(r)$ in the bulk. Lorentz invariance is broken by the non-vanishing temperature, so what is understood as the mass of a meson is frame specific. Our choice follows 
\cite{Mateos:2007vn}, and it is to consider the mass to be defined in the rest frame of the mesons, which is then given by $\omega$ with vanishing three-momentum $k_\mu=0$. The allowed values for $\omega$ are those for which $\chi_1$ remains normalizable near the boundary. By following this procedure we get to a discrete spectrum of frequencies from which we compute the corresponding masses $m_n$, that have a non vanishing value $m_0$ for the ground state, establishing the existence of a mass gap.

\begin{figure}[ht!]
 \centering
 \includegraphics[width=0.42\textwidth]{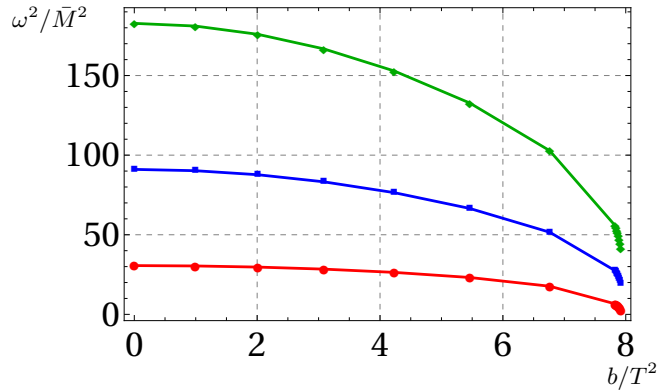}
 \put(-10,-10){$b/T^{2}$}
 \put(-235,125){$\omega^{2}/\bar{M}^{2}$}
\caption{\footnotesize{First three meson masses $\omega^{2}/\bar{M}^{2}$ as a function of $b/T^{2}$ at $\bar{M}/T=1.53$}}
\label{MesonMass}
\end{figure}

To display the impact of the magnetic field on the spectrum, in Fig. 
\ref{MesonMass} we present a plot of its first three masses as a function of $b/T^2$ at $\bar{M}/T=1.53$.

\section{Discussion}

We have exhibited that the presence of an intense magnetic field lowers the temperature at which mesons melt, and demonstrated that this transition can be alternatively driven by applying a magnetic field.

It is important to remember that what we study here is neither the confinement/deconfinement transition, since the existence of a horizon implies that the adjoint degrees of freedom are deconfined, nor the chiral symmetry breaking, since in both phases of our system the expectation value of the chiral condensate is different from zero. It is expected that the transition we study does not coincide with any of the former two, since lattice calculations \cite{Asakawa:2003re,Datta:2003ww,Hatsuda:2005nw} show that the temperature at which the mesons dissociate is indeed higher than that at which deconfinement happens. So, what we explicitly report in this letter is an inverse magnetic catalysis (IMC) for meson dissociation.

IMC for the confinement/deconfinement transition has been studied using holographic methods in \cite{Mamo:2015dea,Rodrigues:2017iqi}, while IMC for chiral symmetry restauration has been reported to happen using different approaches \cite{Bali:2011qj,Agasian:2008tb,Ayala:2015lta,Ayala:2015bgv,Ferreira:2014kpa}, including the holographic Sakai-Sugimoto model at nonzero chemical potential \cite{Preis:2010cq}. We mention this because the geometric reading in \cite{Preis:2010cq} is very similar to ours, since applying a magnetic field makes the D8 and $\bar{\mathrm{D8}}$ branes to bend closer to the horizon causing an early restoration of the chiral symmetry. So, even if the chiral and meson melting IMC are different processes, it seams that their dual gravitational description is of similar nature.

The inset in Fig. \ref{FreePlot} shows the typical pattern of a first order phase transition, making it clear that a more extensive thermodynamic analysis would be enriching, but since this goal is beyond the scope of the present letter, it will be left for future research.

Another result of significant physical relevance is the decrees of the masses in the meson spectrum, including its mass gap, that the magnetic field causes, as can be appreciated in Fig. \ref{MesonMass}. If this effect extrapolates to QCD, it implies that the masses at which some resonances are detected experimentally in non-central collisions are shifted by the influence of the magnetic field produced in them, requiring for some experimental explorations and conclusions to be adapted accordingly.

Fig. \ref{MesonMass} also exhibits a value for $b/T^2$ at which the mass gap vanishes, providing further evidence that the magnetic field dissociates mesons.

Part of our results provide a non-perturbative confirmation of those fund in \cite{Ayala:2018zat}, where a one-loop calculation is done in a linear sigma model to find that the mass of the neutral pion is diminished by the application of a magnetic field, and are in complete agreement with the lattice results in \cite{Bali:2017ian} for the same pion. The spectrum of the $\eta'$ meson was already computed in the holographic context \cite{Zayakin:2008cy}, but as the authors conclude, their calculation, that indicates an increase in the mass with the magnetic field, should be corrected by taking back-reaction effects into account. The same corrections should be applied to \cite{Evans:2010xs}, where the melting transition is studied out of equilibrium. Our results provide such a correction and show that the effect of the field is indeed inverse to the one reported in \cite{Zayakin:2008cy}.

Consistently with previous results, the values that the plots in Fig. \ref{MesonMass} approach at $b/T^2=0$ coincide with those in 
\cite{Mateos:2007vn} for $\bar{M}/T=1.53$.

Among the novel results that we will present in \cite{MagneticMesons} will be the existence of a conformal anomaly in the gauge theory with fundamental degrees of freedom. This anomaly will not only be responsible for the renormalization scheme dependence of some physical quantities, but will also permit to study the enhancement of direct photon production that a magnetic field in its presence has been speculated to cause \cite{Basar:2012bp}.

The holographic study of a hot quark-gluon plasma with massive fundamental degrees of freedom subject to an intense magnetic field is accessible through the supergravity construction presented here, so the lines of research that can be followed using it, and the results of this letter, are numerous. The dynamical nature of many of these processes puts them out of reach for lattice calculations, making the use of our holographic construction particularly appealing. One example is the aforementioned impact of a magnetic field on the luminous spectrum of this system, which we are currently investigating as an extension to the results in 
\cite{Mateos:2007yp,Patino:2012py}. Other examples are the effect of an intense magnetic field on jet quenching or the drag force over quarks on the plasma. It is not our intention to present a comprehensive list of the projects that we are pursuing using the construction that we presented, but we would like to finish by mentioning that we expect many results to derive from our current study.

\begin{acknowledgments}
It is our pleasure to thank Gary Horowitz and Alberto G\"uijosa for very useful discussions, and Francisco Nettel for careful proofreading of this manuscript. We also acknowledge partial financial support from PAPIIT IN113618, UNAM.
\end{acknowledgments}

%
\end{document}